\documentclass[acmsmall]{acmart}

\AtBeginDocument{%
  \providecommand\BibTeX{{%
    \normalfont B\kern-0.5em{\scshape i\kern-0.25em b}\kern-0.8em\TeX}}}

\copyrightyear{2022}
\acmYear{2022}
\setcopyright{usgovmixed}\acmConference[MSIE 2022]{2022 4th International Conference on Management Science and Industrial Engineering (MSIE)}{April 28--30, 2022}{Chiang Mai, Thailand}
\acmBooktitle{2022 4th International Conference on Management Science and Industrial Engineering (MSIE) (MSIE 2022), April 28--30, 2022, Chiang Mai, Thailand}
\acmPrice{15.00}
\acmDOI{10.1145/3535782.3535838}
\acmISBN{978-1-4503-9581-6/22/04}
\usepackage{algorithmic}
\usepackage{tabularx}
\usepackage[ruled,vlined]{algorithm2e}
\usepackage{efbox,graphicx,subfig}
\efboxsetup{linecolor=black,linewidth=.5pt}

\begin{document}

\title{A Deep Learning Approach to Create DNS Amplification Attacks}


\author{Jared Mathews}
\email{jmathew1@citadel.edu}
\authornotemark[1]
\affiliation{%
  \institution{The Citadel}
  \streetaddress{171 Moultrie St}
  \city{Charleston}
  \state{South Carolina}
  \country{USA}
}

\author{Prosenjit Chatterjee}
\affiliation{%
  \institution{The Citadel}
  \streetaddress{171 Moultrie St}
  \city{Charleston}
  \country{USA}}
\email{chatterjee.prosenjit@citadel.edu}

\author{Shankar Banik}
\affiliation{%
  \institution{The Citadel}
  \city{Charleston}
  \country{USA}
  \email{baniks1@citadel.edu}
}

\author{Cory Nance}
\affiliation{%
 \institution{The Citadel}
 \city{Charleston}
 \state{South Carolina}
 \country{USA}
 \email{cnance@citadel.edu}}

\renewcommand{\shortauthors}{Jared Mathews and Prosenjit Chatterjee, et al.}

\begin{abstract}
  In recent years, deep learning has shown itself to be an incredibly valuable tool in cybersecurity as it helps network intrusion detection systems to classify attacks and detect new ones. 
Adversarial learning is the process of utilizing machine learning to 
generate a perturbed set of inputs to then feed to the neural network to misclassify it. Much of the current work in the field of adversarial learning has been conducted in image processing and natural language processing with a wide variety of algorithms. Two algorithms of interest are the Elastic-Net Attack on Deep Neural Networks and TextAttack. In our experiment the EAD and TextAttack algorithms are applied to a Domain Name System amplification classifier. The algorithms are used to generate malicious Distributed Denial of Service adversarial examples to then feed as inputs to the network intrusion detection systems neural network to classify as valid traffic. We show in this work that both image processing and natural language processing adversarial learning algorithms can be applied against a network intrusion detection neural network.
\end{abstract}



\maketitle

\section{Introduction}
Modern network intrusion detection systems (NIDS) have been evolving to utilize the current advancements in artificial intelligence and machine learning, namely deep learning. Deep learning is embedded through the implementation of neural networks. The introduction of deep learning techniques enable the NIDS to detect network threats in a wide range. While
the implementation of neural networks is relatively new in the scope
of classifying network attacks, there has been extensive research
surrounding the use of machine learning classification in other fields
such as image processing and natural language processing (NLP). Researchers have discovered that neural networks are 
particularly vulnerable to adversarial attacks where the data to be 
classified has been perturbed in a way to trick the neural network to receive an incorrect classification \cite{morris2020textattack} \cite{CW} \cite{chen2018ead}. 

However, adversarial attacks have become a real threat to neural networks as shown in 
\cite{Sagduyu}. While these attacks can present significant consequences, these are magnified in the realm of cybersecurity \cite{Sagduyu}. Companies and organizations that rely on NIDS outfitted with neural networks could be vulnerable to serious penetration. Implementations of these NIDS can be vulnerable to these attacks and expose valuable information to malicious actors.

Many attack algorithms have been developed to achieve this goal in image processing such as the Carlini \& Wagner Attack \cite{CW}, Deepfool \cite{DBLP:journals/corr/Moosavi-Dezfooli15}, and fast gradient sign method (FGSM) \cite{DBLP:journals/corr/abs-1806-08970}. One
algorithm that has shown to be superior in this endeavor is the Elastic-Net Attack on Deep Neural Networks (EAD) \cite{chen2018ead}. 
The EAD algorithm has been compared to current Deep Neural Network (DNN) adversarial attack algorithms and has outperformed others with minimal perturbations. These algorithms 
highlight the fragility of modern DNNs. Image classifiers can be effectively fooled with a small amount of perturbation to an image \cite{CW} \cite{chen2018ead} \cite{DBLP:journals/corr/Moosavi-Dezfooli15} \cite{DBLP:journals/corr/abs-1806-08970}.

Adversarial algorithm research in the field of NLP also has yielded many good algorithms. These algorithms are tailored for NLP so they differ greatly from the image processing algorithms. NLP data is discrete whereas images are more continuous, therefore very specialized algorithms are needed to preserve the grammatical structure of the texts being perturbed where with images, the pixels can be perturbed along the color spectrum continuously. The TextAttack algorithm proposed by \cite{morris2020textattack}, is gaining popularity. TextAttack takes in text data and perturbs it by replacing words from a dictionary, deleting characters from words, or even adding characters to a word. Just as in image classification, NLP classifiers are fooled by minimal perturbations to their inputs.

The scope of NIDS utilizing neural networks presents a new attack vector for malicious actors to harm a network. This brings into question the actual resilience of neural networks implemented in a network intrusion context as well as the applicability of these algorithms.

This research and experiment crafts a neural network that effectively can detect DNS amplification attacks and utilize both the EAD algorithm and TextAttack to generate adversarial examples (AEs) that will be misclassified. A  model was created and trained using data from the KDD DDoS-2019 data set 
\cite{8888419} which contains multiple types of DDoS attacks, of which the DNS amplification data \cite{8888419} was used. Figure \ref{DNSAmp} presents a visual representation of the experiment conducted in this paper. Figure \ref{DNSAmp} highlights the experiment's white-box nature with the shared data set. The attacker and the IDS use the same data set for their models. On the IDS side, the victim uses the data set to create a neural network to classify DNS amplification attacks while the attacker utilizes the same data to create AEs based on the victim's model. 
In this project we compare how an image processing and an NLP adversarial algorithm work on network data. We form a neural network to train on KDD DNS data and apply each of these algorithms against it.

\begin{figure*}[!t]
\centering
\efbox{\includegraphics[scale=.25]{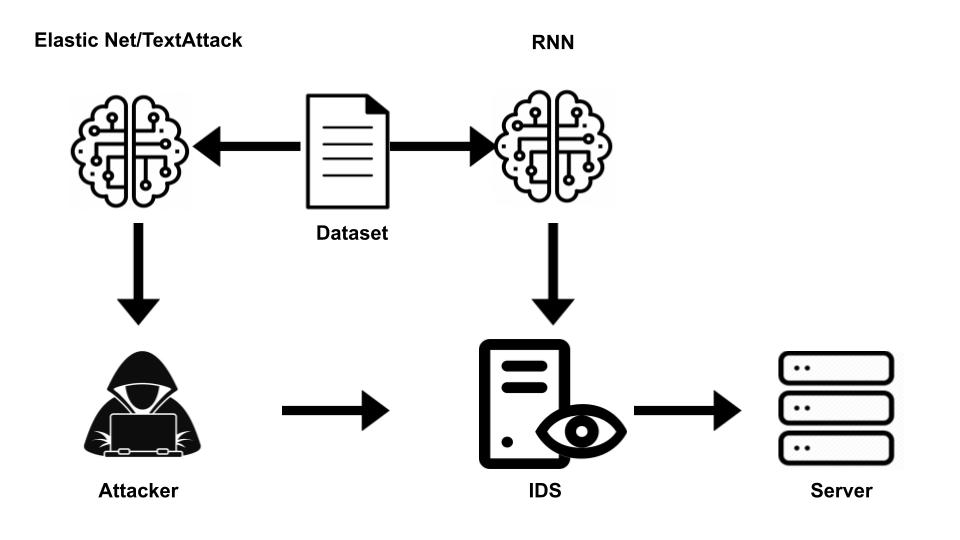}}
\caption{Diagram of Experiment Structure}
\label{threadmodel}
\end{figure*}

\section{Related Work}
Current work in the field of adversarial learning is constantly growing. This technique is being applied in more novel and interesting ways. 
Of the most notable adversarial learning methods is the FGSM, which is an effective method at tricking neural networks \cite{9002856}. There are two methods proposed with FGSM: dodging and impersonation. Dodging makes the classifier mis-classify a normal object, and impersonation makes the classifier think the attacker is a valid user/object. FGSM is a white-box attack. The FGSM attack was used in both modes against a facial recognition neural net with varying perturbation size to determine the most suitable amount. It was shown that with an increase in perturbation, the recognition rate for dodging decreases and the recognition for impersonation increases \cite{9002856}. 

Other examples of such attacks include: the DeepFool method \cite{DBLP:journals/corr/Moosavi-Dezfooli15}, Carlini \& Wagner method \cite{CW}, and the Elastic Net Attack \cite{chen2018ead}. \cite{DBLP:journals/corr/Moosavi-Dezfooli15} implements a new method for generating AEs against convolutional neural nets (CNN). This method fools image classifiers with finer perturbations than used in FGSM. Another well-known method was developed in \cite{CW} that was shown to outperform the DeepFool method. The Carlini \& Wagner attack utilizes a loss function to generate the AEs as well as a distance loss to evaluate the amount of change between the AE and the original image. 

The next method worth mentioning is the Elastic-Net algorithm \cite{chen2018ead}. This method was built off of the Carlini \& Wagner method by treating the process of generating AEs as a regularization problem. This new method was shown to generate AEs with fewer perturbations than the other methods previously mentioned. In addition to generating AEs, these studies also show that their algorithms can be used to make neural networks more robust through adversarial training. Adversarial training is the process of using generated AEs to train a model in order to make it more resilient to attacks.

The EAD algorithm was implemented in \cite{8937669} against deep learning NIDSs and evaluated against other prominent algorithms including: Projected Gradient Descent (PGD), Momentum Iterative Method, DeepFool, and Carlini \& Wagner. These algorithms were tested on a DDoS classifier and each proved to be effective at fooling the model. Though these algorithms were originally designed for image processing, they can be applied to other domains, namely NIDS.

While NLP adversarial algorithms share the same goal with algorithms like EAD, they are very different in their methods. The BERT-based Adversarial Examples (BAE) attack  highlights how NLP adversarial algorithms work \cite{garg2020bae}. Due to the discrete data, an NLP adversarial algorithm needs to work on discrete data as well as be able to preserve grammatical correctness of the perturbed data. BAE works by replacing words and inserting new words into a sentence. Multiple tokens can be inserted or used to mask words, and BAE will choose the most optimal solution. The method provided by BAE shows an effective way to generate AEs for discrete data sets.

Within the field of NLP, attack algorithms also are a growing topic. TextAttack is one tool of particular interest for NLP attacks \cite{morris2020textattack}. TextAttack focuses on text augmentation and implements many documented attacks \cite{alzantot2018generating} \cite{garg2020bae}. Custom attacks can be crafted with TextAttack as well. TextAttack
is open source and provides a robust framework for altering text in multiple ways including individual character operations and synonym replacement. In addition to text augmentation, TextAttack can  train classifiers and run predefined attacks. \cite{morris2020textattack}

\section{Preliminaries}
\subsection{DNS Amplification Attack}
In the field of network attacks, denial of service (DoS) and distributed denial of service (DDoS) attacks are a serious and ongoing threat. An adversary attempts to send numerous data packets through a network targeting a specific server or system with the intent of overwhelming and eventually forcing the device to go offline in a DoS attack. 
The key to a DDoS attack is that there are multiple devices conducting the attack.

\begin{figure*}[!t]
\centering
\efbox{\includegraphics[scale=.25]{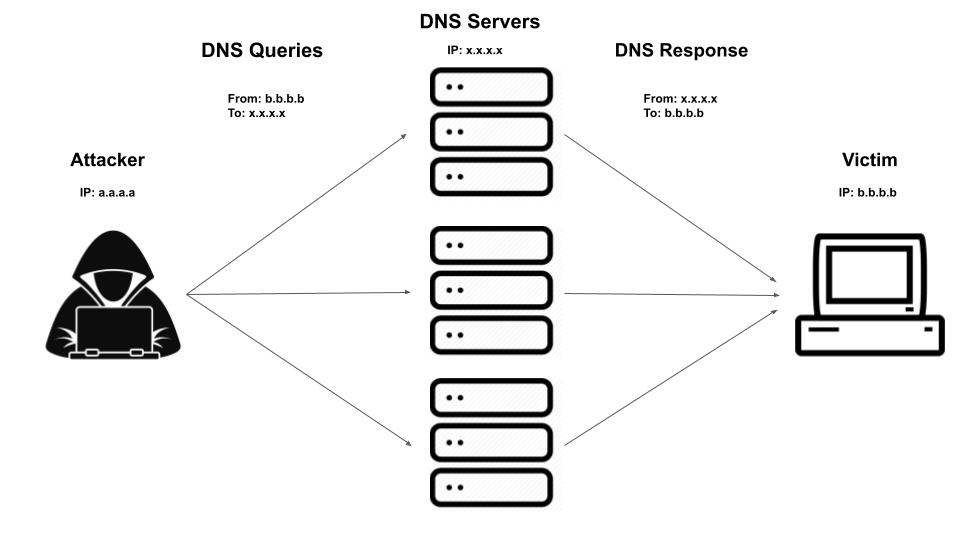}}
\caption{Diagram of DNS Amplification}
\label{DNSAmp}
\end{figure*}

Usually DoS attacks exploit certain protocols used by the victim's system. The Domain Name System (DNS) is a protocol that is particularly interesting for exploitation. DNS is mainly responsible for translating a device IP address into a readable domain name. This protocol presents a unique attack vector for malicious actors. An attacker can easily spoof the victim's IP address and send a DNS name lookup request to an open DNS server which will send the response to the victim, known as an amplification attack. 
a request to a DNS server is significantly smaller than the response. 
DNS amplification attacks are simple and very effective, and due to the open nature of DNS, they are hard to prevent. Figure \ref{DNSAmp} shows the flow of a DNS Amplification attack.

\subsection{Recurrent Neural Networks}

An RNN is a type of neural network that learns sequentially. The data passed into an RNN is that in which the data points are sequential and related through time. The RNN architecture uses the machine learning techniques called long short term memory (LSTM) which acts like a memory block.This mechanism takes the current input, the previous input, and memory of the previous input and decides whether or not to update the memory cell and output the state of the memory cell. This is useful in detecting DoS attacks because the time relationship between packets is the key factor in determining whether or not an attack is occurring.

\subsection{Adversarial Learning}
Adversarial learning is the machine learning process that plays a pivotal role in generating realistic synthetic data. The generated data is used for a robust training of the machine learning embedded model for a consistent classification. These generated inputs are known as adversarial examples (AE). The most notable adversarial learning algorithms are FGSM \cite{DBLP:journals/corr/abs-1806-08970} and Carlini \& Wagner \cite{CW}, which can be either white or black-box. A white-box attack occurs when the data used for training the targeted neural network is known as well as the model structure. However, in a black-box attack, none of the information available in a white-box attack is known. Most of the work in adversarial learning has been done in image processing which presents an opportunity to apply this to intrusion detection systems using deep learning. In \cite{chen2018ead} the authors showed that the EAD algorithm could outperform many of the other commonly used algorithms. The EAD method is a white-box attack as it relies on using the same model to learn how it predicts values and assigns loss in order to modify inputs that will be misclassified.

\section{Methods}
\subsection{Threat Model}
In the EAD attack, the attacker will have a white-box access to the data and model. The EAD algorithm will use the model to determine the most optimal perturbations on the data. For TextAttack, the attacker will have black-box access to the model and will perturb packets without knowledge of the model's output. Figure \ref{threadmodel}

\subsection{Data Set Preprocessing}
\subsubsection{Processes}
To train a neural network, it was necessary to find a robust data set containing enough usable data to feed into the model. The next step was to parse the proper subset of the data to use, select the most optimal features, and shape the data to be passed into the model. KDD has a wide variety of network traffic data for use with NIDS, including the DDoS-2019 data set
\cite{8888419}. The data set contains hundreds of pcap files with millions of packets. The different DDoS attacks are structured by time from two different days where the DNS attacks were between 10:52 - 11:05 (ADT) on the second day. Once the range of pcap files were determined there were as many as 14 million packets with 7,620 DNS packets. Though the whole pcap file was large, a very small portion of it consists of the actual DNS amplification packets needed for training the model. Because of the large disparity between non-attack packets and attack packets, specific slices of the pcap file were selected to utilize all of the DNS packets with normal packets mixed in. The end result was a data set of 16,247 packets with 7,620 DNS packets. The training/test split of the data is as follows: training set containing 10,885 packets with 5,105 DNS packets, and a testing set of 5,362 packets with 2,515 DNS packets. The pcap files mix the DDoS packets with the normal traffic packets which adds a bit of complexity to parsing and labeling the training data. CICFlowMeter was used in \cite{8888419} to parse the pcap features into csv format. The CICFlowMeter returns a csv file with features obtained from packet flows parsed from the pcap file. In our experiment a custom pcap parsing method using the python library scapy was used to extract full packets and determine which features were best. The experiment used the entire packet for direct perturbations in the adversarial algorithms.

Much of the data contained in packets are in plain English and not numerical. Neural networks need the data fed to them to be numerical. Common methods used in NLP for training on non-numerical data were used to achieve this. Tokenization is one of the more prevalent methods of converting non-numerical data to numerical. Tokenization maps each unique object to an index in an array and allows all the pcap data to be converted into integer values as seen in Algorithm \ref{alg:token}. The values were then normalized in the array from -0.5 to +0.5 to make the data easier to process for the model. This range is also the range used from \cite{chen2018ead}. The method for normalization is shown in Algorithm \ref{alg:norm}.

\begin{algorithm}[]
\SetAlgoLined
\textbf{Input: }
1D Matrix $a_{m}$ containing packet features, Matrix $B_{mn}$ with all non tokenized values from each packet\\
\textbf{Output: } 
Vector $\overrightarrow{b}$ with features tokenized\\
\For{i = 0 to $m$}{
    \eIf{$\overrightarrow{a_i} \notin B_i$}{
        $B_i \cup \overrightarrow{a_i}$\\
        $\overrightarrow{b_i} \coloneqq n \: where \: B_{in} = \overrightarrow{a_i}$ \\
    }{
        $\overrightarrow{b_i} \coloneqq n \: where \: B_{in} = \overrightarrow{a_i}$ \\
        
    }
}
\caption{Tokenization Algorithm}
 \label{alg:token}
\end{algorithm}

\begin{algorithm}[]
\SetAlgoLined
\textbf{Input: }
Padded Tensor $S_{xyz}$, 
lower value $low$, higher value: $high$\\
\textbf{Output: } 1D Array of Normalized Values $b$\\
${a} \coloneqq \bigcup_{i = 0}^x \bigcup_{j = 0}^y \bigcup_{k = 0}^z S_{ijk}$\\
$|{a}| = x*y*z$\\

${b} \bigcup_{l=0}^{|{a}|} \frac{({a}_l - min({a}))*(high - low)}{max({a}) - min({a})} + low$\\
 \caption{Normalization Algorithm}
 \label{alg:norm}
\end{algorithm}

The features that lend themselves to least bias needed to be determined after parsing the raw pcap data from the KDD files.
\cite{8888419} determined for the DNS amplification attack that max packet length was the most important feature for classifying DNS amplification attacks. 
To accomplish this, a genetic algorithm \cite{info11050243} was implemented to determine which features from a given csv file are less likely to produce training bias. Originally the packets were split into 42 features (TCP and UDP), and these were narrowed to 15 features after passing the data through the genetic algorithm. The features used were the destination address, source address, IP packet length, IP id,IP flags, IP chksum, DNS id, DNS ra, source port, destination port, seq, ack, dataofs, chksum, urgptr.

To capture the nature of a DDoS attack for the training process, the data needed to be re-structured to reflect the temporal nature of this attack for the labeling process. The frequency of DNS packets is the core of the attack so this needed to be quantified. The packet relationships are another important aspect of these attacks. To label the data in a meaningful way the initial set of packets were split into subsets based on their temporal relationship with each other. Algorithm \ref{alg:packet} was used to generate every subset of the original set that contained all packets within 30 seconds of each other. This allowed the KDD data set \cite{8888419} to be extended and generate packet flows for each packet in the set. \cite{8888419} found that the most important feature in labeling DNS amplification attacks is the packet length. It was determined that the average max length of DNS attack packet was 1378.80 bytes. The maximum DNS packet length was 465 bytes with the majority of malicious responses above 100 bytes. Each packet flow was labeled as an attack if within the 30 seconds it contained at least 3 DNS packets of size greater than 100. This left us with a ragged tensor, which means the set of packet flows is not symmetrical where each packet flow contains a unique amount of packets. Tensors in a ragged shape are more difficult to work with than with a symmetrical shape. Another method used was tensor padding, which is shown in Algorithm \ref{alg:padd}. Padding works by appending null or neutral values to the end of sets of data in order to fit them all to the same size. The number of packets in a flow was the portion of the data that was not symmetrical among all of them. The process takes a three dimensional tensor 
$T_{ijk}$ where \textit{k} represents a packet, and \textit{j} represents a flow of packets and is unique among all \textit{i}. \textit{k} is a set of packet features with cardinality 15. The largest matrix of packets was found and then each of smaller size matrices had the correct amount of neutral packets appended, where a neutral packet is defined as one that doesn't affect the labeling of the data. The data was labeled according to the predetermined metric for an attack. The labels were merely a 0 for no attack and a 1 for an attack.

\begin{algorithm}[]
\SetAlgoLined
\textbf{Input: }
Matrix of all packets \textit{p} in pcap file: $P_{nm}$ 
where \textit{n} is the number of packets and \textit{m} is the number of features selected from packet, \textit{t(p)} is a function to retrieve the timestamp from packet, and time threshold $t$\\
\textbf{Output} 
Three dimensional tensor $T_{njm}$\\ 
 \For{i = 0 to n
 }{
    $low = i$\;
    $high = n$\;
    $high_o = high$\;
    \While{$low \leq high$
    }{
        $middle = low + \frac{high - low}{2}$\;
        \eIf{$t(P_{low  m}) - t(P_{high  m}) < t$
        }{
            $g = 0$\;
            \For{x = middle to $high_o + 1$
            }{
                \eIf{$t(P_{low  m} - t(P_{xm}) < t$}{
                    $g = x + 1$\;                
                }{
                    $g = x$\;
                }
                }
                $T_{njm} \cup \bigcup_{k = low}^g P_{km}$\;
        }{
            $high_o = high + 1$\;
            $high = middle - 1$\;
        }
    }
 }
 
 \caption{Packet Structuring Algorithm}
 \label{alg:packet}
\end{algorithm}

\begin{algorithm}[]
\SetAlgoLined
\textbf{Input: }
Ragged Three Dimensional tensor: $T_{njm}$,
 max $j$ size: $n$,
 and packet for padding $p_m$\\
\textbf{Output} 
Padded Tensor $PT_{njm}$\\

\For{i = 0 to n}{
  \If{$T_{jm} < T_{nm}$}{
    \For{$n - j$}{
        $T_{jm} \cup \bigcup_{n = 0}^{n-j} p_m$\\
    }
  }
}
 \caption{Tensor Padding Algorithm}
 \label{alg:padd}
\end{algorithm}

\subsubsection{Metrics}
To evaluate the effectiveness of these algorithms, time metrics were observed for each algorithm proposed in this paper. The results of these metrics can be found in Table \ref{metrics}. Each algorithm was evaluated for five different pcap files containing 500, 1,000, 5,000, 10,000, and 100,000 packets. The average time to process each packet is approximately 71.839 ms. Realistically all the algorithms will not be used together. The more interesting comparison is between the padding algorithms and the ragged algorithms. 
We observed that the padded arrays typically takes 0.69 ms slower to normalize and the actual padding algorithm takes 1.648 ms less than converting the array to a ragged tensorflow tensor.
Overall the average time to process one packet with padding takes 62.169 ms and the ragged algorithms take 63.127 ms. This difference of .958 ms is relatively small when evaluating just one packet, but for every 1,000 packets, this magnifies to approximately one second longer. This becomes too inefficient when processing hundreds of thousands or millions of packets.

\subsection{Model}

\begin{figure*}[!t]
\centering
\efbox{\includegraphics[trim={0 3cm 0 2cm},clip, scale=.55]{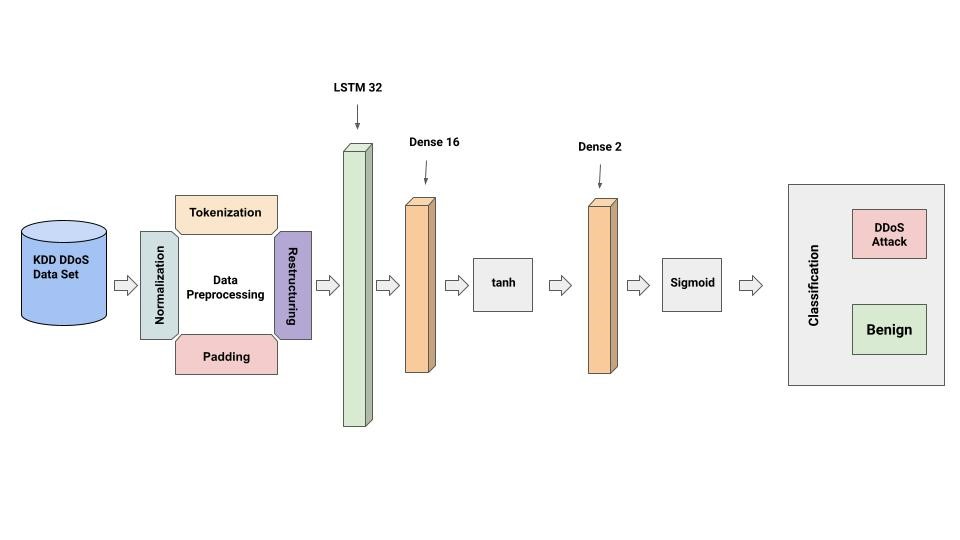}}
\caption{Diagram of RNN Model}
\label{model}
\end{figure*}

An RNN was used for the classifier using an LSTM layer, due to the temporal nature of DDoS attacks and pcap data. The model consists of an LSTM layer and two dense layers using a tanh, and a sigmoid activation. Classification accuracy ranged from 98\% to 99\% with the final model. The main structure of data used was the padded tensor with 15 features mentioned above. The ragged tensor and the padded tensor were passed into the model for evaluation. Both ragged and padded tensors were formed from the same data set and were passed into the same models with the same batch size and number of epochs. After testing both of these structures of tensor, it was observed that the differences in the accuracy achieved was almost negligible as the two produced nearly identical predictions of the testing data. At most, the difference between the two was 1\%. Figure \ref{model} shows a visual representation of the flow through the deep neural network.

\begin{table*}
  \caption{Different Stages of Preprocessing Algorithms Timed in Milliseconds}
  \label{metrics}
  \begin{tabular}{p{0.1\linewidth}p{0.1\linewidth}p{0.1\linewidth}p{0.1\linewidth}p{0.1\linewidth}p{0.1\linewidth}cc}
    \toprule
    Algorithm & 500 packets & 1,000 packets & 5,000 packets & 10,000 packets & 100,000 packets & Packets/ms & ms/packet \\
    \midrule
    Loading pcap & 771.64 & 2730.98 & 61132.22 & 233337.90 & 22701422.19 & 0.23 & 53.37\\ 
    Format and Label & 41.51 & 85.51 & 446.57 & 891.66 & 8842.05 & 11.49 & 0.087\\
    Padding & 42.00 & 84.51 & 332.05 & 807.14 & 9675.30 & 12.30 & 0.082\\
    Normalize Padded & 3670.64 & 7471.30 & 43390.11 & 88800.57 & 1075670.57 & 0.11 & 8.63\\
    Normalize Ragged & 3788.66 & 7886.88 & 40914.17 & 81999.88 & 787955.13 & 0.13 & 7.94\\
    Conversion to Ragged & 1304.30 & 1961.19 & 6998.75 & 13896.63 & 128880.77 & 0.62 & 1.73\\
    
  \bottomrule
\end{tabular}
\end{table*}

\subsection{EAD Algorithm}
Once an effective model to classify DNS amplification attacks was formed, 
the EAD algorithm was adapted to it. The goal was to preserve the functionality of the algorithm while allowing it to work on three-dimensional packet tensors. Originally the data passed into the 
algorithm was a four-dimensional tensor 
$I_{ijjk}$ where \textit{i} 
represents the number of images, \textit{j} represents the \textit{x} and 
\textit{y} axis of a square image which is set as 28, and \textit{k} represents 
the number of channels which was set as 1. This tensor can be significantly smaller than the ones generated from this project. The size of the image tensor 
can be shown as 784\textit{n} where \textit{n} is the number of images. The packet 
tensor size can be shown as 21\textit{l}\textit{m} where \textit{l} is the number of 
packet flows and \textit{m} is the number of packets in a flow. This tensor grows to be very cumbersome for larger selections of pcap files as \textit{l} is 16,247 in this experiment and \textit{m} is typically in the thousands. For the attack the authors use an "iterative  shrinkage-thresholding  algorithm (ISTA)" \cite{chen2018ead} to shrink certain pixel values given their difference to the original values are greater than a defined threshold. This method is similar to the Carlini \& Wagner algorithm but more efficient \cite{CW}. The attack takes batches of images or packet flows and compares these to a lower bound 
\begin{equation}\label{eq1}
    L_{njm} = \bigcup_{i = 0}^{i= n}l_{jm}
\end{equation}

of the same shape as the input tensor and where $l_{jm}$ is a matrix with all values of 0. The upper bound is also a tensor

\begin{equation}\label{eq2}
U_{njm} = \bigcup_{i = 0}^{i= n}u_{jm}
\end{equation}

of the same shape as the input and $u_{jm}$ is a matrix with all values of $10^{10}$.

Scores are assigned to the bounds and the inputs and evaluated to determine which is most optimal. From here the best attack is selected and used on the model.

\subsection{TextAttack Algorithm}
After evaluating EAD on the model, TextAttack needed to be adapted to perturb packet data. TextAttack lends itself to packet data more than the EAD algorithm as the packet data is inherently discrete and non-continuous. The TextAttack python library
is very robust, open-source, and easily adapted to this research \cite{morris2020textattack}. The python library contains multiple modes of operation for augmenting data including the following: CLAREAugmenter, CheckListAugmenter, EasyDataAugmenter, CharSwapAugmenter, EmbeddingAugmenter, and WordNetAugmenter. Of these modes, ClareAugmenter and CharSwapAugmenter were used. The CLAREAugmenter replaces, inserts, and merges text with a pre-trained masked language model. CharSwapAugmenter substitutes, deletes, inserts, and swaps adjacent characters. TextAttack can take in single words or lists of words to perturb and can be chained. To apply TextAttack to the packet data, it first needed to be determined which of the packet features were mutable. Not all features of the packets are text so a method for applying TextAttack to integer values also was needed. To accomplish this, an algorithm to convert each individual digit in an integer to its respective alphabetical value which is between 'a' and 'i'. The CharSwapAugmenter then was used to modify the converted number through multiple iterations and then convert the word back into an integer. The same data used with the EAD algorithm was also used with TextAttack as both were capable of perturbing the same tensor structures. In contrast to the EAD algorithm, TextAttack is able to perturb the packet data while maintaining legitimate packet features making it a more practical attack.

\section{Results}
\begin{figure*}[t]
\centering

\subfloat[Confusion Matrix from RNN]{%
  \includegraphics[width=5cm]{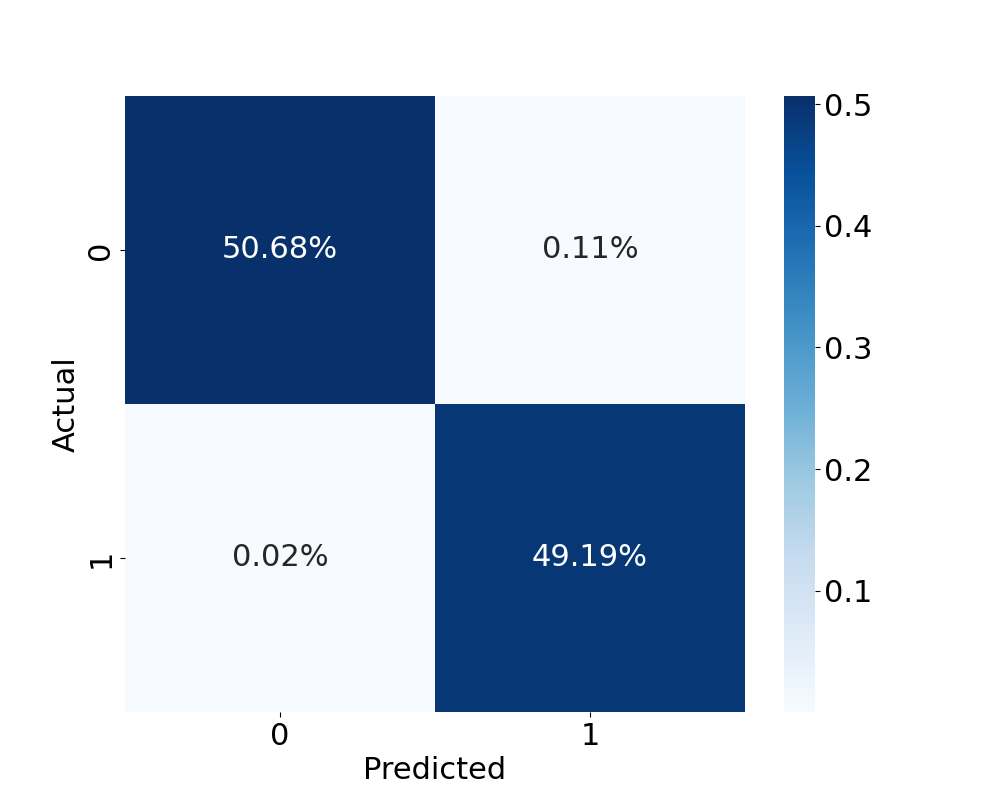}%
  \label{RNN-CM}%
}\qquad
\subfloat[Confusion Matrix from TextAttack]{%
  \includegraphics[width=5cm]{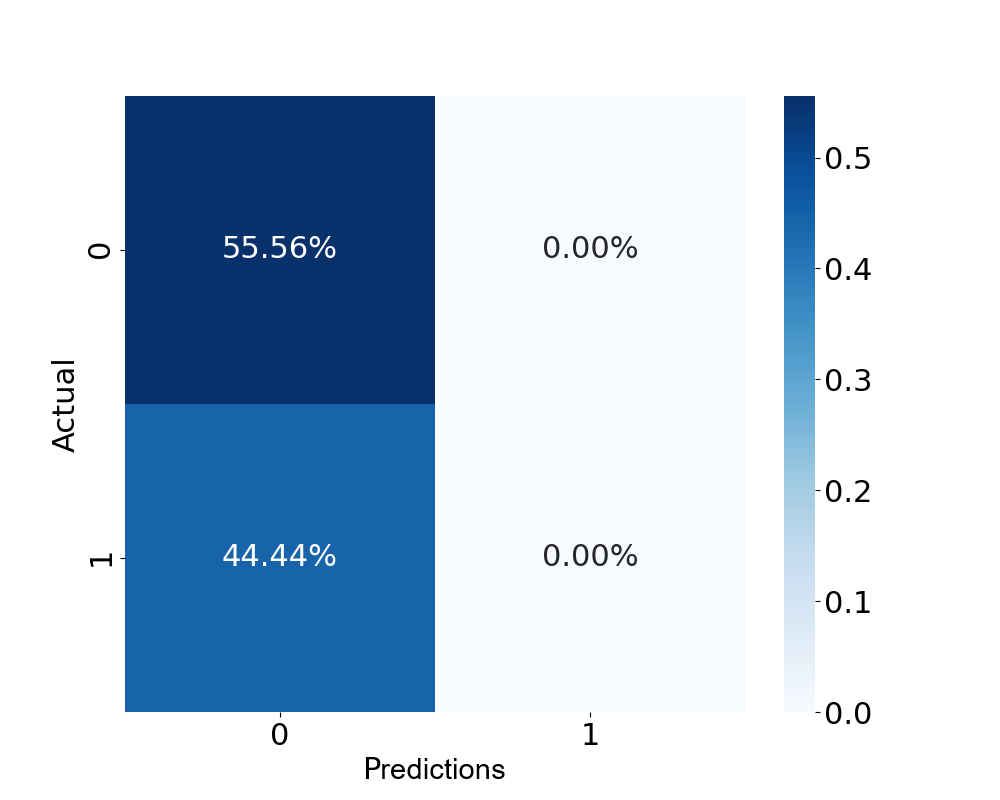}%
  \label{TACM}%
}\qquad
\subfloat[Confusion Matrix from EAD]{%
  \includegraphics[width=5cm]{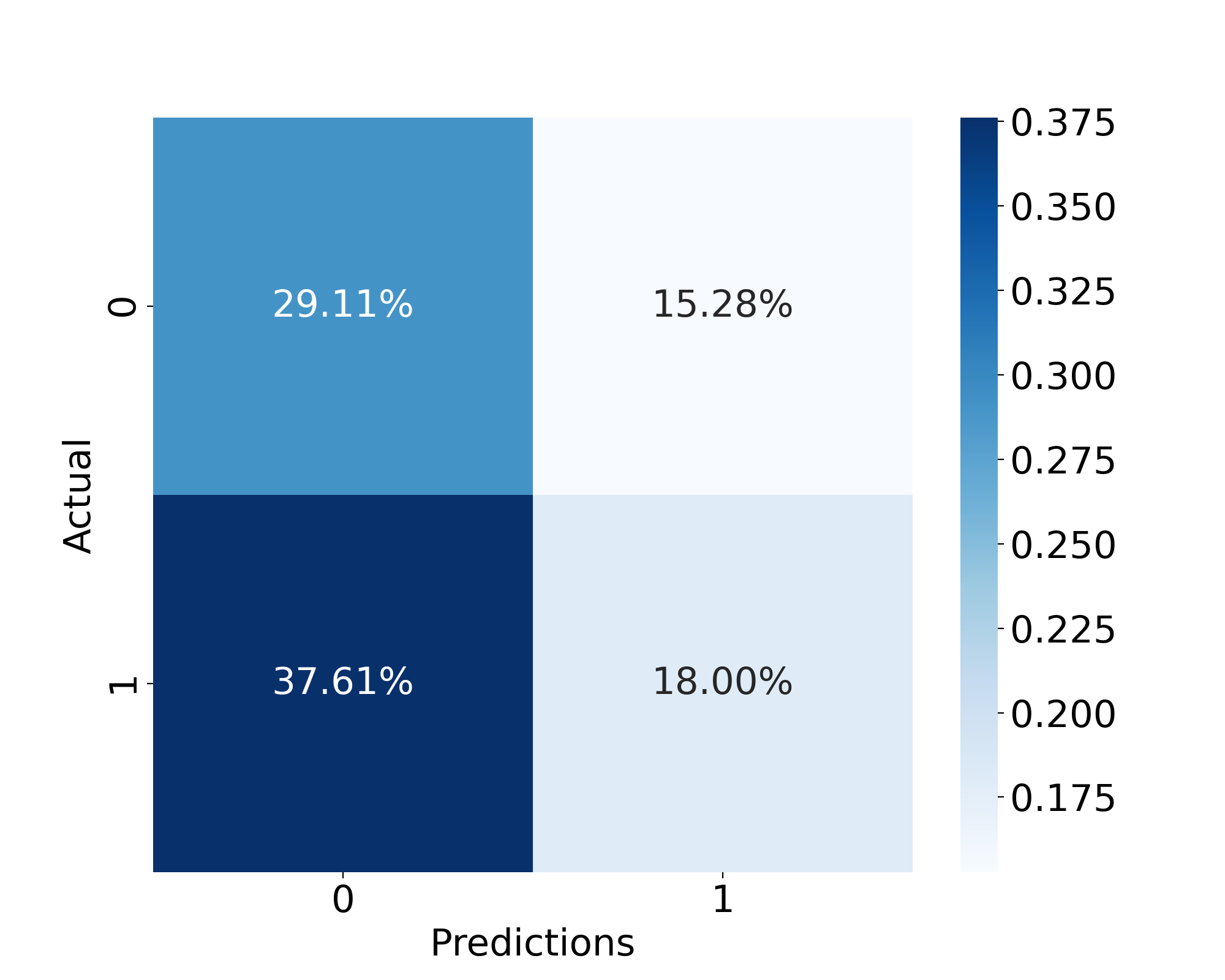}%
  \label{EADCM}%
}

\caption{Simulation results}
\end{figure*}

To evaluate the results of the EAD and TextAttack experiment, two metrics were used: average perturbation, and attack success rate. A baseline of the model was taken before conducting the attacks as seen in Figure \ref{RNN-CM}.

To calculate the average perturbation of the AEs created by each algorithm the distances between unperturbed packets and perturbed packets was taken. To calculate this percentage for each pair of packets, the percent difference was evaluated for each feature in the packets and then averaged as displayed in Algorithm \ref{alg:perturb}. The results can be seen in Table \ref{ASR}. There is a huge disparity between the two algorithms with regards to average percent perturbation. The reason for EADs high perturbation rate is due to the fact it consistently perturbed the packet features all to $0$ as a part of its optimization, resulting in a 200\% distance. The TextAttack creates a more normal distance as each individual packet features individually were slightly perturbed by single character insertions or deletions.

\begin{algorithm}[]
\SetAlgoLined
\textbf{Input: }
Original packet $p$, perturbed packet $p'$\\
\textbf{Output} 
Percent difference\\

$n = \lvert p \rvert =  	\lvert p' \rvert$; \\
$\frac{1}{n}$  $\sum_{i=0}^{n} \frac{\lvert p_i - p'_i\rvert}{\frac{p_i + p'_i}{2}};$ \\
 \caption{Average Percent Perturbation Algorithm}
 \label{alg:perturb}
\end{algorithm}

The results for the attack success rate is shown in Figures \ref{TACM} and \ref{EADCM}. With TextAttack, the AEs were able to succeed in deceiving the classifier with regard to attacks while also allowing benign packet flows to remain benign. The EAD confusion matrix also shows a similar outcome with slightly more variance. Both of the attacks had a high percentage of false positives which was the goal of the attacks. They also had a high percentage of true positives which show that they can preserve benign packets without converting them into classifiable attacks. The percent of attacks that were classified as benign was used to calculate the attack success rate. Table \ref{ASR} shows the results of the attacks success rates. The results show that both algorithms are capable of producing quality AEs that are capable of fooling a DNS amplification classifier.

\begin{table}
  \caption{EAD vs TextAttack Perturbation Percentage and Success Rate}
  \label{ASR}
  \begin{tabular}{ccl}
    \toprule
    Attack & Success Rate & Percent Perturbed \\
    \midrule
    EAD & 67.63& 200.00\\
    TextAttack & 100.00 & 24.95 \\
  \bottomrule
\end{tabular}
\end{table}

\section{Conclusion and Future Work}
In our research, a RNN was formed to train on DNS amplification data. The EAD and TextAttack algorithms were applied to this model to deceive it. The two algorithms were evaluated and compared based on their performance with network traffic data and how well they preserved the nature of the data. Results show that it is possible and relatively easy to deceive a machine learning NIDS, reaffirming the notion that these deep learning algorithms are quite susceptible to adversarial learning. It is possible to adapt adversarial algorithms created for image processing or NLP to a network classifier. While these algorithms are capable of perturbing network traffic data, they don't necessarily craft realistic packets, leading to future development of a new adversarial algorithm that is purely intended for network traffic classifiers. We found that the TextAttack algorithm can generate AEs with 100\% chance of deception against the model. The AEs from the EAD algorithm had a 67.63\% chance to deceive the model. The perturbation rate of the TextAttack algorithm was 24.95\% where the EAD algorithm perturbed the packets by 200\%.

In the future, the goal is to create a new adversarial attack designed specifically for attacks on network traffic classifiers and implement defenses through adversarial learning and training distillation. 

Our next future goal is to apply the work done on DNS amplification classifiers to IoT DDoS attacks. In particular, focusing on DDoS attacks against the constrained application protocol (CoAP) used by many IoT devices. This would include creating a data set similar to that from KDD of CoAP traffic including malicious packets. This would then be applied to a real world simulation of an IoT environment using a NIDS.

\bibliographystyle{ACM-Reference-Format}
\bibliography{references}


\begin{thebibliography}{12}


\ifx \showCODEN    \undefined \def \showCODEN     #1{\unskip}     \fi
\ifx \showDOI      \undefined \def \showDOI       #1{#1}\fi
\ifx \showISBNx    \undefined \def \showISBNx     #1{\unskip}     \fi
\ifx \showISBNxiii \undefined \def \showISBNxiii  #1{\unskip}     \fi
\ifx \showISSN     \undefined \def \showISSN      #1{\unskip}     \fi
\ifx \showLCCN     \undefined \def \showLCCN      #1{\unskip}     \fi
\ifx \shownote     \undefined \def \shownote      #1{#1}          \fi
\ifx \showarticletitle \undefined \def \showarticletitle #1{#1}   \fi
\ifx \showURL      \undefined \def \showURL       {\relax}        \fi
\providecommand\bibfield[2]{#2}
\providecommand\bibinfo[2]{#2}
\providecommand\natexlab[1]{#1}
\providecommand\showeprint[2][]{arXiv:#2}

\bibitem[Abusnaina et~al\mbox{.}(2019)]%
        {8937669}
\bibfield{author}{\bibinfo{person}{Ahmed Abusnaina}, \bibinfo{person}{Aminollah
  Khormali}, \bibinfo{person}{DaeHun Nyang}, \bibinfo{person}{Murat Yuksel},
  {and} \bibinfo{person}{Aziz Mohaisen}.} \bibinfo{year}{2019}\natexlab{}.
\newblock \showarticletitle{Examining the Robustness of Learning-Based DDoS
  Detection in Software Defined Networks}. In \bibinfo{booktitle}{\emph{2019
  IEEE Conference on Dependable and Secure Computing (DSC)}}.
  \bibinfo{pages}{1--8}.
\newblock
\urldef\tempurl%
\url{https://doi.org/10.1109/DSC47296.2019.8937669}
\showDOI{\tempurl}


\bibitem[Alzantot et~al\mbox{.}(2018)]%
        {alzantot2018generating}
\bibfield{author}{\bibinfo{person}{Moustafa Alzantot}, \bibinfo{person}{Yash
  Sharma}, \bibinfo{person}{Ahmed Elgohary}, \bibinfo{person}{Bo-Jhang Ho},
  \bibinfo{person}{Mani Srivastava}, {and} \bibinfo{person}{Kai-Wei Chang}.}
  \bibinfo{year}{2018}\natexlab{}.
\newblock \bibinfo{title}{Generating Natural Language Adversarial Examples}.
\newblock
\newblock
\showeprint[arxiv]{1804.07998}~[cs.CL]


\bibitem[Carlini and Wagner(2016)]%
        {CW}
\bibfield{author}{\bibinfo{person}{Nicholas Carlini} {and}
  \bibinfo{person}{David~A. Wagner}.} \bibinfo{year}{2016}\natexlab{}.
\newblock \showarticletitle{Towards Evaluating the Robustness of Neural
  Networks}.
\newblock \bibinfo{journal}{\emph{CoRR}}  \bibinfo{volume}{abs/1608.04644}
  (\bibinfo{year}{2016}).
\newblock
\showeprint[arxiv]{1608.04644}
\urldef\tempurl%
\url{http://arxiv.org/abs/1608.04644}
\showURL{%
\tempurl}


\bibitem[Chen et~al\mbox{.}(2018)]%
        {chen2018ead}
\bibfield{author}{\bibinfo{person}{Pin-Yu Chen}, \bibinfo{person}{Yash Sharma},
  \bibinfo{person}{Huan Zhang}, \bibinfo{person}{Jinfeng Yi}, {and}
  \bibinfo{person}{Cho-Jui Hsieh}.} \bibinfo{year}{2018}\natexlab{}.
\newblock \bibinfo{title}{EAD: Elastic-Net Attacks to Deep Neural Networks via
  Adversarial Examples}.
\newblock
\newblock
\showeprint[arxiv]{1709.04114}~[stat.ML]


\bibitem[Garg and Ramakrishnan(2020)]%
        {garg2020bae}
\bibfield{author}{\bibinfo{person}{Siddhant Garg} {and}
  \bibinfo{person}{Goutham Ramakrishnan}.} \bibinfo{year}{2020}\natexlab{}.
\newblock \bibinfo{title}{BAE: BERT-based Adversarial Examples for Text
  Classification}.
\newblock
\newblock
\showeprint[arxiv]{2004.01970}~[cs.CL]


\bibitem[Liu et~al\mbox{.}(2019)]%
        {9002856}
\bibfield{author}{\bibinfo{person}{Yujie Liu}, \bibinfo{person}{Shuai Mao},
  \bibinfo{person}{Xiang Mei}, \bibinfo{person}{Tao Yang}, {and}
  \bibinfo{person}{Xuran Zhao}.} \bibinfo{year}{2019}\natexlab{}.
\newblock \showarticletitle{Sensitivity of Adversarial Perturbation in Fast
  Gradient Sign Method}. In \bibinfo{booktitle}{\emph{2019 IEEE Symposium
  Series on Computational Intelligence (SSCI)}}. \bibinfo{pages}{433--436}.
\newblock
\urldef\tempurl%
\url{https://doi.org/10.1109/SSCI44817.2019.9002856}
\showDOI{\tempurl}


\bibitem[Milton(2018)]%
        {DBLP:journals/corr/abs-1806-08970}
\bibfield{author}{\bibinfo{person}{Md~Ashraful~Alam Milton}.}
  \bibinfo{year}{2018}\natexlab{}.
\newblock \showarticletitle{Evaluation of Momentum Diverse Input Iterative Fast
  Gradient Sign Method {(M-DI2-FGSM)} Based Attack Method on {MCS} 2018
  Adversarial Attacks on Black Box Face Recognition System}.
\newblock \bibinfo{journal}{\emph{CoRR}}  \bibinfo{volume}{abs/1806.08970}
  (\bibinfo{year}{2018}).
\newblock
\showeprint[arxiv]{1806.08970}
\urldef\tempurl%
\url{http://arxiv.org/abs/1806.08970}
\showURL{%
\tempurl}


\bibitem[Moosavi{-}Dezfooli et~al\mbox{.}(2015)]%
        {DBLP:journals/corr/Moosavi-Dezfooli15}
\bibfield{author}{\bibinfo{person}{Seyed{-}Mohsen Moosavi{-}Dezfooli},
  \bibinfo{person}{Alhussein Fawzi}, {and} \bibinfo{person}{Pascal Frossard}.}
  \bibinfo{year}{2015}\natexlab{}.
\newblock \showarticletitle{DeepFool: a simple and accurate method to fool deep
  neural networks}.
\newblock \bibinfo{journal}{\emph{CoRR}}  \bibinfo{volume}{abs/1511.04599}
  (\bibinfo{year}{2015}).
\newblock
\showeprint[arxiv]{1511.04599}
\urldef\tempurl%
\url{http://arxiv.org/abs/1511.04599}
\showURL{%
\tempurl}


\bibitem[Morris et~al\mbox{.}(2020)]%
        {morris2020textattack}
\bibfield{author}{\bibinfo{person}{John~X. Morris}, \bibinfo{person}{Eli
  Lifland}, \bibinfo{person}{Jin~Yong Yoo}, \bibinfo{person}{Jake Grigsby},
  \bibinfo{person}{Di Jin}, {and} \bibinfo{person}{Yanjun Qi}.}
  \bibinfo{year}{2020}\natexlab{}.
\newblock \bibinfo{title}{TextAttack: A Framework for Adversarial Attacks, Data
  Augmentation, and Adversarial Training in NLP}.
\newblock
\newblock
\showeprint[arxiv]{2005.05909}~[cs.CL]


\bibitem[Muhuri et~al\mbox{.}(2020)]%
        {info11050243}
\bibfield{author}{\bibinfo{person}{Pramita~Sree Muhuri},
  \bibinfo{person}{Prosenjit Chatterjee}, \bibinfo{person}{Xiaohong Yuan},
  \bibinfo{person}{Kaushik Roy}, {and} \bibinfo{person}{Albert Esterline}.}
  \bibinfo{year}{2020}\natexlab{}.
\newblock \showarticletitle{Using a Long Short-Term Memory Recurrent Neural
  Network (LSTM-RNN) to Classify Network Attacks}.
\newblock \bibinfo{journal}{\emph{Information}} \bibinfo{volume}{11},
  \bibinfo{number}{5} (\bibinfo{year}{2020}).
\newblock
\showISSN{2078-2489}
\urldef\tempurl%
\url{https://doi.org/10.3390/info11050243}
\showDOI{\tempurl}


\bibitem[Sagduyu et~al\mbox{.}(2019)]%
        {Sagduyu}
\bibfield{author}{\bibinfo{person}{Yalin~E. Sagduyu}, \bibinfo{person}{Yi Shi},
  {and} \bibinfo{person}{Tugba Erpek}.} \bibinfo{year}{2019}\natexlab{}.
\newblock \showarticletitle{IoT Network Security from the Perspective of
  Adversarial Deep Learning}.
\newblock \bibinfo{journal}{\emph{CoRR}}  \bibinfo{volume}{abs/1906.00076}
  (\bibinfo{year}{2019}).
\newblock
\showeprint[arxiv]{1906.00076}
\urldef\tempurl%
\url{http://arxiv.org/abs/1906.00076}
\showURL{%
\tempurl}


\bibitem[Sharafaldin et~al\mbox{.}(2019)]%
        {8888419}
\bibfield{author}{\bibinfo{person}{Iman Sharafaldin},
  \bibinfo{person}{Arash~Habibi Lashkari}, \bibinfo{person}{Saqib Hakak}, {and}
  \bibinfo{person}{Ali~A. Ghorbani}.} \bibinfo{year}{2019}\natexlab{}.
\newblock \showarticletitle{Developing Realistic Distributed Denial of Service
  (DDoS) Attack Dataset and Taxonomy}. In \bibinfo{booktitle}{\emph{2019
  International Carnahan Conference on Security Technology (ICCST)}}.
  \bibinfo{pages}{1--8}.
\newblock
\urldef\tempurl%
\url{https://doi.org/10.1109/CCST.2019.8888419}
\showDOI{\tempurl}


\end{thebibliography}

\end{document}